\documentclass[conference]{IEEEtran}
\IEEEoverridecommandlockouts
\usepackage{amsmath,amssymb,amsfonts}
\usepackage{algorithmic}
\usepackage{graphicx}
\usepackage{textcomp}
\usepackage{xcolor}
\usepackage{subfig}
\usepackage{multirow}
\usepackage[numbers,sort&compress]{natbib}
\def\BibTeX{{\rm B\kern-.05em{\sc i\kern-.025em b}\kern-.08em
    T\kern-.1667em\lower.7ex\hbox{E}\kern-.125emX}}
\begin{document}

\title{A Robust Anti-noise Scheme for RF Fingerprint Identification\\

\thanks{This work was supported in part by the National Key Research and Development Program of China (2022YFB4300300), National Natural Science Foundation of China under Grant 62171120, Jiangsu Provincial Key Laboratory of Network and Information Security No. BM2003201, Guangdong Key Research and Development Program under Grant 2020B0303010001, Chongqing Natural Science Joint Fund Project under Grant No. CSTB2023NSCQ-LZX0121, and Purple Mountain Laboratories for Network and Communication Security.}
}

\author{\IEEEauthorblockN{1\textsuperscript{st} Junxian Shi}
\IEEEauthorblockA{\textit{School of Cyber Science and} \\
\textit{Engineering, Southeast University}\\
Nanjing, China \\
junxianshi@seu.edu.cn}\and
\IEEEauthorblockN{2\textsuperscript{th} Linning Peng}
\IEEEauthorblockA{\textit{School of Cyber Science and} \\
\textit{Engineering, Southeast University}\\
Nanjing, China \\
pengln@seu.edu.cn}\and
\IEEEauthorblockN{3\textsuperscript{nd} Wentao Jing}
\IEEEauthorblockA{\textit{School of Cyber Science and} \\
\textit{Engineering, Southeast University}\\
Nanjing, China \\
wentaojing@seu.edu.cn}
\and
\IEEEauthorblockN{4\textsuperscript{th} Lingnan Xie}
\IEEEauthorblockA{\textit{School of Cyber Science and} \\
\textit{Engineering, Southeast University}\\
Nanjing, China \\
lnxie@seu.edu.cn}\and
\IEEEauthorblockN{5\textsuperscript{rd} Haichuan Peng}
\IEEEauthorblockA{\textit{School of Cyber Science and} \\
\textit{Engineering, Southeast University}\\
Nanjing, China \\
penghc@seu.edu.cn}\and
\IEEEauthorblockN{6\textsuperscript{th} Aiqun Hu}
\IEEEauthorblockA{\textit{School of Information Science and} \\
\textit{Engineering, Southeast University}\\
Nanjing, China \\
aqhu@seu.edu.cn}
}

\maketitle

\begin{abstract}
    Radio frequency (RF) fingerprint technology is utilized for wireless device identification, extensively employed in the internet of things (IoT). The operating environment for IoT devices is challenging, with pervasive noise and distortion on the signals which blur the feature space of RF fingerprints. Consequently, the model accuracy obtained through training at high signal-to-noise ratio (SNR) scenarios decreases with the low SNR of the received signals in testing. To solve the noise domain adaptation problem, an anti-noise scheme is proposed to enhance identification accuracy of RF fingerprint at varying SNRs. The squared cross power spectral density (SCPSD) features are first proposed to obtain the same RF fingerprint representation. Subsequently, the specific effect of noise on SCPSD is theoretically derived and the rationality of the scheme is demonstrated through simulation experiments. Finally, 60 off-the-shelf ZigBee devices are employed to evaluate the performance of the anti-noise algorithm. The experimental results show that employing the random subspace k-nearest neighbors (RSKNN) classifier not only effectively classifies devices with multi-cluster feature, but combined with the anti-noise scheme can significantly improve the accuracy by approximately 46\% for SNRs not less than 5~dB.

\end{abstract}


\section{Introduction}
The commercialization of fifth-generation (5G) communication technology and advancements in sixth-generation (6G) research have significantly propelled the growth of the Internet of Things (IoT) industry \cite{10486565}. However, the inherent openness of wireless channels exposes numerous devices to threats such as denial-of-service (DoS) attacks, man-in-the-middle (MITM) attacks, and unauthorized access \cite{8524450}. Additionally, the portability of wireless IoT devices often results in limited computation and storage capabilities, hindering the implementation of high-performance security protocols \cite{10457524}.

Radio frequency (RF) fingerprint technology, characterized by its uniqueness, long-term stability, and resistance to cloning, shows significant promise in identifying diverse wireless devices \cite{10487783,10436740,10105269}. Derived from hardware variations, RF fingerprints serve as a reliable identity marker, similar to biometric fingerprints for individuals. However, since RF fingerprints are inherently derived from the transmitted signal, noise interference and channel fading during transmission pose significant challenges \cite{10417054,10184130,10188589}. In low signal-to-noise ratio (SNR) scenarios, noise can significantly disrupt signal amplitude patterns, blurring device-specific characteristics.

To address noise interference in narrowband signals like ZigBee \cite{8360937}, LoRa \cite{10476864}, and Bluetooth \cite{10323345}, this paper focuses on enhancing RF fingerprint extraction. For narrowband signals, channel fading can often be approximated as flat fading, making noise interference the primary impediment to accurate RF fingerprint extraction. By employing noise reduction techniques specifically tailored for narrowband signals, we aim to improve the reliability and accuracy of RF fingerprint identification in noisy environments.

In the previous study of noise resistance \cite{10139783}, the cross power spectral density (CPSD) serves as the RF fingerprint with high device identification accuracy at the same SNR, which had not yet provided mutual corroboration under different SNRs. Therefore, we redesigned the RF fingerprint extraction method based on CPSD, termed squared CPSD (SCPSD). We illustrate the effects of noise on SCPSD at different SNRs through formula derivation and experimental validation. An anti-noise algorithm is finally proposed to address the issue where models trained at high SNRs perform poorly at low SNRs, known as the noise domain adaptation problem. The main contributions are summarized as follows:
\begin{itemize}[\IEEEsetlabelwidth{Z}]
  \item
  The noise-containing form of the SCPSD is explicitly represented in the formula, where a detailed analysis is conducted to investigate the specific impact of noise on each term. The analysis and processing algorithms for noise in different terms are validated by simulation experiments. A set of flowcharts for the anti-noise algorithms is developed.

  \item
  The random subspace k-nearest neighbors (RSKNN) classifier is proposed to achieve high accuracy for the case of spatially non-unique distribution of RF fingerprints extracted from real devices. The effectiveness of the RSKNN classifier on classification task of multi-cluster feature is illustrated by a comparative analysis with other classifiers.

  \item
  The application of the proposed anti-noise algorithm to 60 off-the-shelf ZigBee devices significantly improves classification accuracy by approximately 46\% compared to its performance without the algorithm, especially when SNR is at least 5 dB.

\end{itemize}

\section{Related Work}
The studies on RF fingerprint against noise can be classified into three groups: those based on statistical properties, those based on various transformations, and those based on machine learning techniques. 

A representative idea of methods based on statistical properties is superposition. Xing et al. \cite{8469002} achieved an improvement in SNR by averaging the accumulation of multiple signal frames with the same symbol. The 900 signal frames are accumulated with a SNR of -15 dB, resulting in a final classification accuracy of 98.5\%. However, this approach relies on the assumption that the superimposed signal frames are from the same device and lacks practicality and generalizability. Wang et al. \cite{9766175} rotated the constellation diagrams under different modulation methods to enable the superimposition of different symbols with less length of training data. It also combines with smoothing filtering to further weaken the noise effect, obtaining an overall accuracy of more than 93.32\%.

The core principle of noise reduction methods using transformations is that the useful signal and the noise are separated by transformations. Subsequently, the noisy component could be suppressed or eliminated, thereby reducing noise. Louis et al. \cite{WOS:000808852300001} employed singular value decomposition to decompose the received signal into signal and noise subspaces. They then used a selection algorithm to choose eligible eigenvalues for the signal subspace, resulting in noise reduction. Xie et al. \cite{8470925} utilised the discrete wavelet transform to decompose the received signal into a series of components. Only coefficients exceeding a predetermined threshold were considered valid signals and retained, while others were removed. In addition, a smoothed signal with suppressed noise was obtained through wavelet reconstruction.

The most common method based on machine learning techniques is data augmentation \cite{9247526,10304522}. These methods enhance classifier performance by incorporating features at various SNRs during training, which can enable the classifier to learn the feature distribution across different noise levels, ultimately improving its performance on the testing set. The numerous existing studies have evaluated the performance of the proposed method by introducing artificial white Gaussian noise (AWGN) to the received signal at high SNRs, simulating the received signal at low SNRs. Such an approach lacks both experimental validation and in-depth analysis of the underlying principles of data augmentation, primarily due to the non-interpretability of neural networks.

\section{RF Fingerprint Extraction Algorithm}
The relationship in the time domain between the preprocessed received signal $y$ and the ideal baseband signal $x$ can be expressed mathematically as
\begin{equation}
\label{eq_1}
    y=x\otimes r\!f\!f + g,
\end{equation}
where $\!r\!f\!f$ denotes RF fingerprint of transmitter, $g$ denotes Gaussian noise and $\otimes$ denotes circular convolution. For narrowband signals, the channel fading is flat fading, so the channel response is negligible.

According to the Wiener-Hinchin theorem, the CPSD and the cross correlation function can be converted to each other by Fourier transform and inverse Fourier transform. Due to the influence of $g$, there is a change in the energy of $y$, which is unfavorable for the subsequent extraction of RF fingerprint. Therefore, the CPSD of energy normalized $\bar{y}$ can be expressed as
\begin{equation}
\label{eq_3}
\begin{aligned}
  &\!\!\!\!\!\!\!C\!P\!S\!D\!=\!abs(\!f\!ft(\bar{y})\!\!\cdot\!\! conj(\!f\!ft(conj(x))))\\
   &\!\!\!\!\!\!\!=\!abs(\!f\!ft(\frac{1}{P}y)\!\!\cdot\!\! conj(\!f\!ft(conj(x))))\\
   &\!\!\!\!\!\!\!=\!\!\frac{1}{P}abs(\!f\!ft(x\otimes r\!f\!f + g)\!\!\cdot\!\! conj(\!f\!ft(conj(x))))\\
   &\!\!\!\!\!\!\!=\!\!\frac{1}{P}abs((X\!\!\cdot\!\! R\!F\!F + G)\!\!\cdot\!\! X')\\
   &\!\!\!\!\!\!\!=\!\!\frac{1}{P}abs(X \!\!\cdot\!\! R\!F\!F \!\!\cdot\!\! X' + G \!\!\cdot\!\! X' )\\
   &\!\!\!\!\!\!\!=\!\!\frac{\left ( \! \left \| \! X\!\!\cdot\!\! R\!F\!F \!\!\cdot\!\! X\!' \!\right \| ^{\!2} \!\!+\!\! \left \| \! G \!\!\cdot\!\! X\!' \! \right \| ^{\!2} \!\!+\!\! 2\!\! \left \|\! X \!\!\cdot\!\! R\!F\!F \!\!\cdot\!\! X\!' \!\right \| \! \left \| \!G \!\!\cdot\!\! X\!'\! \right \|\! cos\theta \! \right )\!\!^\frac{1}{2} }{P}
\end{aligned},
\end{equation}
where $\!f\!ft$ denotes fast Fourier transform, $abs$ denotes modulo operation, $conj$ denotes conjugate operation, $X, R\!F\!F, G$ correspond to the Fourier transform results of $x, r\!f\!f, g$ respectively and $\theta$ is the angle between $X \!\!\cdot\!\! R\!F\!F \!\!\cdot\!\! X'$ and $G \!\!\cdot\!\! X'$. The last equals sign uses the cosine theorem. The energy normalization parameter P in (\ref{eq_3}) can be expressed as
\begin{equation}
\label{eq_4}
\begin{aligned}
    P&=\sqrt{\frac{\sum_{i=1}^{L_N} \left \|  y_i \right \| ^{2}}{L_N} }=\sqrt{\frac{\sum_{i=1}^{L_N} \left \|  x_i^{r\!f\!f} + g_i \right \| ^{2}}{L_N} } \\
    &=\sqrt{\frac{\sum_{i=1}^{L_N} \left \| x_i^{r\!f\!f} \right \| ^{2} +  \left \| g_i \right \| ^{2} + 2\left \| x_i^{r\!f\!f} \right \| \left \| g_i \right \| cos\alpha_i}{L_N} } 
\end{aligned},
\end{equation}
where $y_i$ denotes the i-th sampling point in $y$, $L_N$ denotes the length of $y$ and $\alpha_i$ denotes the angle between $x_i^{r\!f\!f}$ and $g_i$.

Combining (\ref{eq_3}) and (\ref{eq_4}) shows that the numerator and denominator both contain square root operations, which is detrimental to the derivation of the formulas. Therefore, the root sign is removed and a new representation of the device characteristics is proposed as SCPSD, which is represented as
\begin{equation}
\label{eq_5}
\begin{aligned}
    &S\!-\!C\!P\!S\!D = C\!P\!S\!D ^{2}\\
    &\!\!=\!\frac{\left \| \! X\!\! \cdot\!\! R\!F\!F \!\!\cdot\!\! X' \right \| ^{\!2}\!\! +\!\! \left \| G \!\!\cdot\!\! X' \right \| ^{\!2} + 2 \left \| X\!\! \cdot \!\!R\!F\!F \!\!\cdot\!\! X' \right \| \left \| G \!\!\cdot\!\! X' \right \| cos\theta}{\frac{1}{L_N}\sum_{i=1}^{L_N} \left \| x_i^{r\!f\!f} \right \| ^{\!2} +  \left \| g_i \right \| ^{\!2} + 2\left \| x_i^{r\!f\!f} \right \| \left \| g_i \right \| cos\alpha_i}\\
    &\!\!=\!\frac{\left \|\! X \! \right \| ^{\!2} \!\left \|\! R\!F\!F\! \right \| ^{\!2}\! \left \|\! X'\! \right \| ^{\!2} \!\!+\!\! \left \|\! G \!\right \| ^{\!2}\! \left \|\! X'\! \right \| ^{\!2}\!\! +\!\! 2 \!\left \|\! X \!\right \|\! \left \| \!R\!F\!F\!\right \|\! \left \|\! X' \!\right \| \!\left \|\! G\! \right \| \!\left \|\! X'\! \right \| \!cos\theta}{\frac{1}{L_N}\sum_{i=1}^{L_N} \left \| x_i^{r\!f\!f} \right \| ^{\!2} +  \left \| g_i \right \| ^{\!2} + 2\left \| x_i^{r\!f\!f} \right \| \left \| g_i \right \| cos\alpha_i}\\
    &\!\!=\!\frac{\left \|\! X \! \right \| ^{\!2} \!\left \|\! R\!F\!F\! \right \| ^{\!2}\! \left \|\! X'\! \right \| ^{\!2} \!\!+\!\! \left \|\! G \!\right \| ^{\!2}\! \left \|\! X'\! \right \| ^{\!2}\!\! +\!\! 2\! \left \| \!X\! \right \|\! \left \|\! R\!F\!F\!\right \|\!  \left \| \!G \!\right \|\! \left \|\! X' \!\right \|^{\!2}\!cos\theta }{\frac{1}{L_N}\sum_{i=1}^{L_N} \left \| x_i^{r\!f\!f} \right \| ^{\!2} +  \left \| g_i \right \| ^{2} + 2\left \| x_i^{r\!f\!f} \right \| \left \| g_i \right \| cos\alpha_i}
\end{aligned}.
\end{equation}

\section{Anti-noise Processing}
As can be seen from (\ref{eq_5}), both the numerator and denominator contain noise terms, which have different effects. In the absence of noise, the SCPSD results contain only the RF fingerprint. To achieve noise immunity, it is necessary to quantify the effect of noise on SCPSD separately.

\subsection{Normalized Energy item Processing}
The signal-to-noise ratio is commonly used in communication systems to describe the energy relationship between the effective signal and the noise in the received signal, defined as
\begin{equation}
\label{eq_6}
    S\!N\!R=10\log_{10}{\frac{\sum_{i=1}^{L_N}\left \| x_i \right \| ^{2}}{\sum_{i=1}^{L_N}\left \| g_i \right \| ^{2}}}.
\end{equation}
Then the noise power in the received signal can be expressed as
\begin{equation}
\label{eq_7}
    \sum_{i=1}^{L_N}\left \| g_i \right \| ^{2}=10^{-0.1 \cdot S\!N\!R}\sum_{i=1}^{L_N}\left \| x_i \right \| ^{2}.
\end{equation}
Bringing (\ref{eq_7}) into the energy normalization term $P^2$ yields
\begin{equation}
\label{eq_8}
\!\!\!\!   \frac{\sum_{i=1}^{L_N} \!\! \left \| x_i^{r\!f\!f} \right \| ^{2} \!\! + \!\!  10^{-\frac{S\!N\!R}{10}}\sum_{i=1}^{L_N} \!\!\left \| x_i \right \| ^{2} \!\!+ \!\!2 \left \| x_i^{r\!f\!f} \right \| \left \| g_i \right \| cos\alpha_i}{L_N}.
\end{equation}
From (\ref{eq_8}), it can be analyzed that $P^2$ can be approximated as an exponential function of the independent variable as $S\!N\!R$, denoted as
\begin{equation}
\label{eq_9}
    P^2=a \cdot 10^{-b \cdot S\!N\!R}+c.
\end{equation}
Comparing (\ref{eq_8}) and (\ref{eq_9}) yields that $b$ is approximately equal to 0.1, while $c$ is approximately equal to 1. The result for $b$ is obvious, while $c$ for further explanation. $c$ is theoretically equal to $\sum_{i=1}^{L_N} \left \| x_i^{r\!f\!f} \right \| ^{2}$, and in the absence of noise, the normalized energy of the signal with RF fingerprint is 1. 

\begin{figure}[htbp]
    \centering
        \subfloat{
        \includegraphics[width=0.23\textwidth]{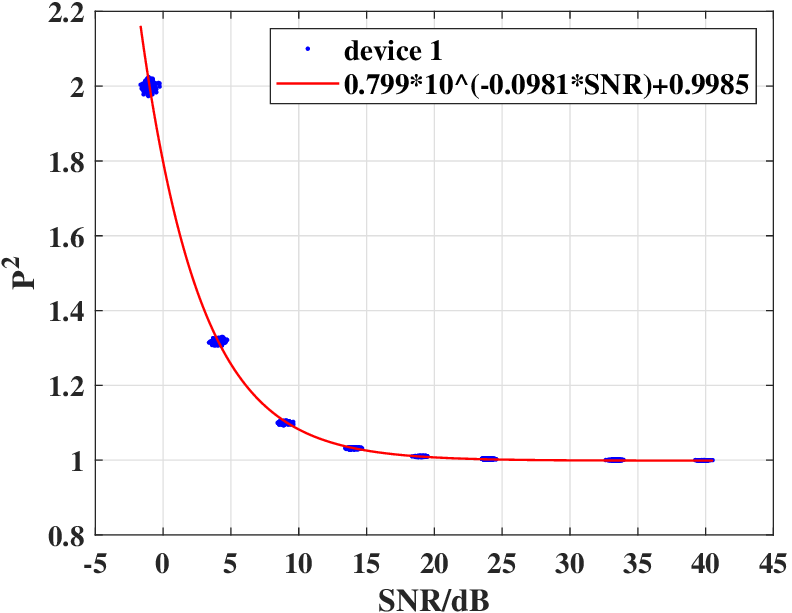}}       
        \subfloat{
        \includegraphics[width=0.23\textwidth]{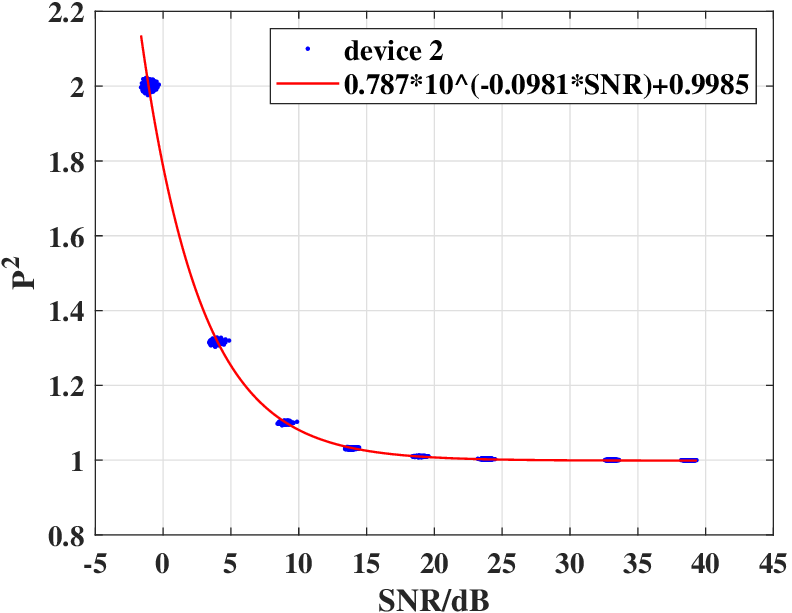}}
        \caption{Variation curves of signal energy with SNR for device 1 and device 2.}
    \label{fig1}
\end{figure}

To demonstrate the validity of the aboved elaboration, 60 off-the-shelf ZigBee devices were selected for the experiment. 300 frames of signal were captured from each device, and the SNR at receiver was greater than 40 dB. The SNR of each signal frame was varied from 0 dB to 30 dB in steps of 5 dB by adding AWGN. The signal energy was calculated under different SNRs and the results are shown in Fig.~\ref{fig1}, taking device 1 and device 2 as examples.

\begin{figure}[htbp]
    \centering
    \includegraphics[width=0.23\textwidth]{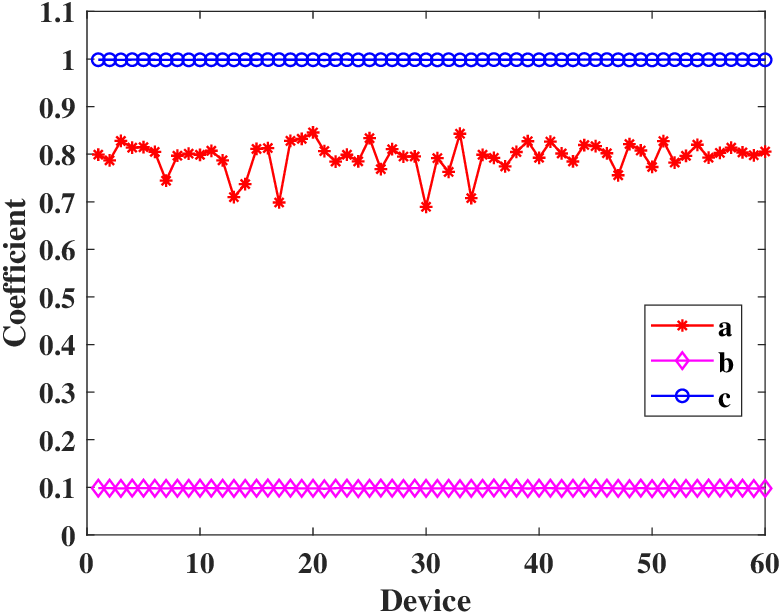}
\caption{Coefficient curves for 60 ZigBee devices.}
\label{fig2}
\end{figure}

Fig.~\ref{fig1} shows that the results of curve fitting the 2400 sample points for each device are consistent with the assumptions of (\ref{eq_9}). The values of $b$ and $c$ are as expected when errors are tolerable. To better demonstrate this trend, the coefficient values for all 60 devices are shown in Fig.~\ref{fig2}.

As seen in Fig.~\ref{fig2}, the coefficients $b$ and $c$ for the 60 devices are almost very close to 0.1 and 1, respectively, in line with the inference. For coefficient $a$, it is important to note that each device carries a different RF fingerprint, resulting in a variation of $\sum_{i=1}^{L_N}\left \| x_i \right \| ^{2}$ for each device when $\sum_{i=1}^{L_N} \left \| x_i^{r\!f\!f} \right \| ^{2}=1$. By obtaining the coefficients corresponding to each device in the training phase, it is possible to compensate the change in normalized energy caused by noise for each device in the testing phase.

\subsection{RF fingerprint item processing}
The SNR is calculated and then the energy-normalized estimation $\hat{P^2}$ is obtained to compensate for the SCPSD. In the case of a better compensation, the remaining RF fingerprint item can be expressed as

\begin{equation}
\label{eq_10}
\begin{aligned}
    SCPSD \cdot \hat{P^2} =& \left \| X  \right \| ^{2} \left \| R\!F\!F \right \| ^{2} \left \| X' \right \| ^{2} + \left \| G \right \| ^{2} \left \| X' \right \| ^{2} \\
    &+ 2 \left \| X \right \| \left \| R\!F\!F\right \|  \left \| G \right \| cos\theta \left \| X' \right \|^{2}
\end{aligned},
\end{equation}
In real-world scenarios, signal frames vary in function and length. Therefore, the RF fingerprint of a signal frame is extracted using a single symbol as a unit. According to (\ref{eq_10}), the RF fingerprint that can be extracted from a symbol is expressed as
\begin{equation}
\label{eq_11}
\begin{aligned}
\!\!\!    S_j^{R\!F\!F}=& \left \| X_{S_j}  \right \| ^{2} \left \| R\!F\!F_{S_j} \right \| ^{2} \left \| X_{S_j}' \right \| ^{2} + \left \| G_{S_j} \right \| ^{2} \left \| X_{S_j}' \right \| ^{2} \\
    &+ 2 \left \| X_{S_j} \right \| \left \| R\!F\!F_{S_j} \right \|  \left \| G_{S_j} \right \| cos\theta_{S_j} \left \| X_{S_j}' \right \|^{2}
\end{aligned},
\end{equation}
where $S_j^{R\!F\!F}$ denotes the RF fingerprint extracted from the j-th symbol, $X_{S_j}$ denotes the spectrum of the j-th symbol and the subscripts of other symbols have the same meaning. 

Next, each term in (\ref{eq_11}) is analyzed. In terms of the definition of RF fingerprint, the RF fingerprint extracted from either a symbol or a frame of a signal is the same and does not vary with symbol. Furthermore, since ZigBee symbols satisfy the cyclic shift characteristic, $R\!F\!F_{S_j}$ is the same for all symbols. Multiplied by the RF fingerprint, the item $\left \| X_{S_j}  \right \| ^{2} \left \| R\!F\!F_{S_j} \right \| ^{2} \left \| X_{S_j}' \right \| ^{2}$ is then ideally a fixed sequence of constants. However, in practice this term may fluctuate slightly between symbols due to device instability. Therefore, it is considered to average the results extracted from multiple symbols to obtain a stable RF fingerprint representation. The RF fingerprint $F^{R\!F\!F}$ of a frame of signal can eventually be represented as

\begin{equation}
\label{eq_12}
\begin{aligned}
    &F^{R\!F\!F}=\frac{1}{N_{s\!y\!m}}\sum_{j=1}^{N_{s\!y\!m}}S_j^{R\!F\!F}\\
    &=\frac{1}{N_{\!s\!y\!m}}\!\!\!\sum_{j=1}^{N_{s\!y\!m}}\! ( \left \| X_{S_{\!j}}  \right \| ^{\!2} \!\left \| R\!F\!F_{S_{\!j}} \right \| ^{\!2} \!\left \| X_{S_{\!j}}' \right \| ^{\!2}\!\! +\!\! \left \| G_{S_{\!j}} \right \| ^{\!2}\! \left \| X_{S_{\!j}}' \right \| ^{\!2} \\
    & \quad+ 2 \left \| X_{S_{\!j}} \right \| \left \| R\!F\!F_{S_{\!j}} \right \|  \left \| G_{S_{\!j}} \right \| cos\theta_{S_{\!j}} \left \| X_{S_{\!j}}' \right \|^{\!2}  ) \\
    &=\left \| X  \right \| ^{\!2} \left \| R\!F\!F \right \| ^{\!2} \left \| X' \right \| ^{\!2} + \left \| X' \right \| ^{\!2} \frac{\sum_{j=1}^{N_{s\!y\!m}} \left \| G_{S_{\!j}} \right \| ^{\!2}}{N_{s\!y\!m}}\\
    & \quad+2\left \| X \right \| \left \| R\!F\!F \right \|\left \| X' \right \| ^{\!2}\frac{\sum_{j=1}^{N_{s\!y\!m}} \left \| G_{S_{\!j}} \right \| cos\theta_{S_{\!j}}}{N_{s\!y\!m}}
\end{aligned},
\end{equation}
where $N_{s\!y\!m}$ denotes the total number of symbols in a signal frame. Each symbol in the received signal can be determined by a decoding algorithm and is therefore a constant sequence. For the second term with noise, it can be further derived as

\begin{equation}
\label{eq_13}
\begin{aligned}
    &\left \| X' \right \| ^{2} \frac{\sum_{j=1}^{N_{s\!y\!m}} \left \| G_{S_j} \right \| ^{2}}{N_{s\!y\!m}}\\
    &=\left \| X' \right \| ^{2} \frac{\sum_{j=1}^{N_{s\!y\!m}} \left \| X_{S_j}^{R\!F\!F} \right \| ^{2} \cdot 10^{-0.1 \cdot S\!N\!R}}{N_{s\!y\!m}} \\
    &=\left \| X' \right \| ^{2} \frac{\sum_{j=1}^{N_{s\!y\!m}} E_{s\!y\!m} \cdot 10^{-0.1 \cdot S\!N\!R}}{N_{s\!y\!m}}\\
    &=\left \| X' \right \| ^{2} \frac{\sum_{j=1}^{N_{s\!y\!m}} L_s \cdot 10^{-0.1 \cdot S\!N\!R}}{N_{s\!y\!m}}\\
    &=\left \| X' \right \| ^{2} \cdot L_s \cdot 10^{-0.1 \cdot S\!N\!R}
\end{aligned},
\end{equation}
where $X_{S_j}^{R\!F\!F}$ denotes the j-th symbol in the signal with RF fingerprint, $E_{s\!y\!m}$ denotes the energy of a symbol and $L_s$ denotes the length of a symbol. The first equals sign in (\ref{eq_13}) utilizes the SNR definition in the frequency domain. After performing energy normalization, the energy of a symbol is numerically equal to the symbol length. From (\ref{eq_13}), it can be concluded that noise has a positive effect on the RF fingerprint in proportion to the SNR, shifting the overall SCPSD curve upward. The $cos\theta$ in the last remaining item with noise can be positive or negative, so that it will cause the magnitude of the SCPSD curve to fluctuate up and down. However, when the number of symbols is quite large, the mean value of $\sum_{j=1}^{N_{s\!y\!m}} \left \| G_{S_j} \right \| cos\theta_{S_j}$ is 0.

\begin{figure}[htbp]
    \centering
        \subfloat[Overall view]{
        \label{fig3_1}
        \includegraphics[width=0.23\textwidth]{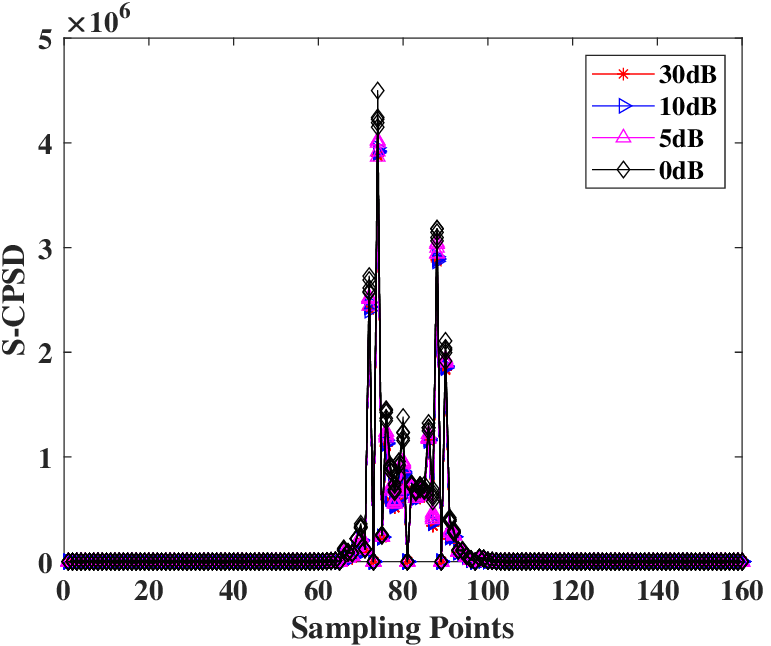}}
        \subfloat[Local enlargement]{
        \label{fig3_2}
        \includegraphics[width=0.23\textwidth]{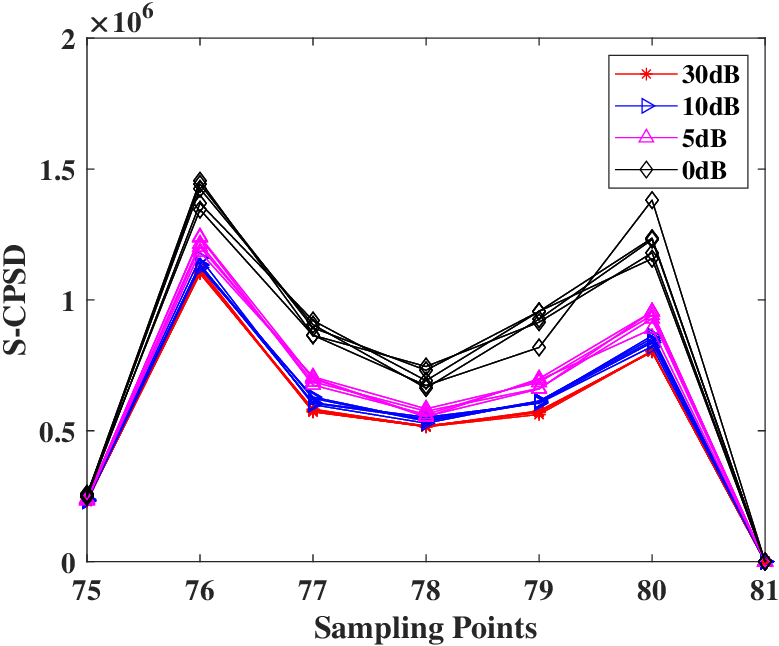}}
        \caption{SCPSDs of device 1 at different SNRs without energy normalization.}
    \label{fig3}
\end{figure}

To visualize the variation of each term in (\ref{eq_12}), the SCPSD is extracted after adding AWGN to the received signals with SNRs greater than 40 dB. However, note that the energy normalization operation is not performed, assuming that compensation for the energy normalization term has already been applied. Using device 1 as an example, the SCPSD magnitude variation with SNRs is shown in Fig.~\ref{fig3}.

\begin{figure}[htbp]
    \centering
        \subfloat[Device 1]{
        \label{fig4_1}
        \includegraphics[width=0.23\textwidth]{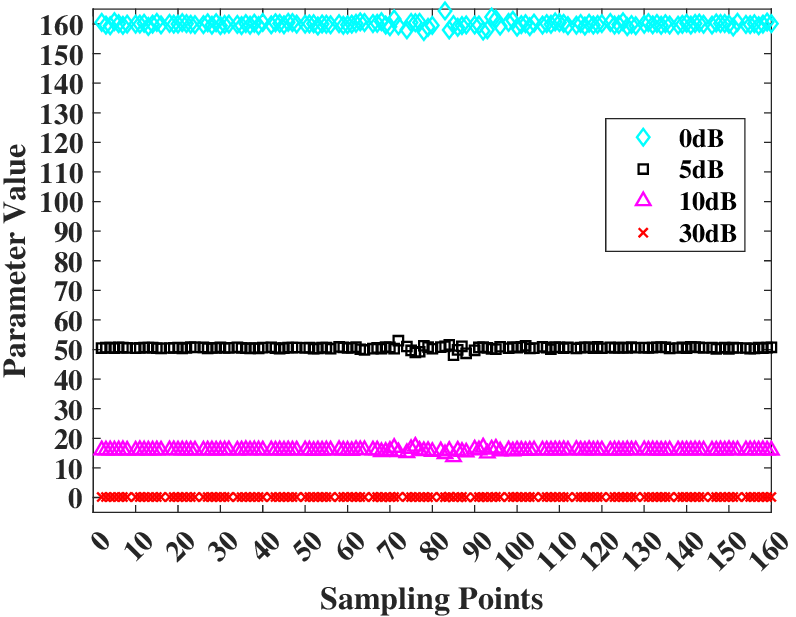}}
        \subfloat[Device 2]{
        \label{fig4_2}
        \includegraphics[width=0.23\textwidth]{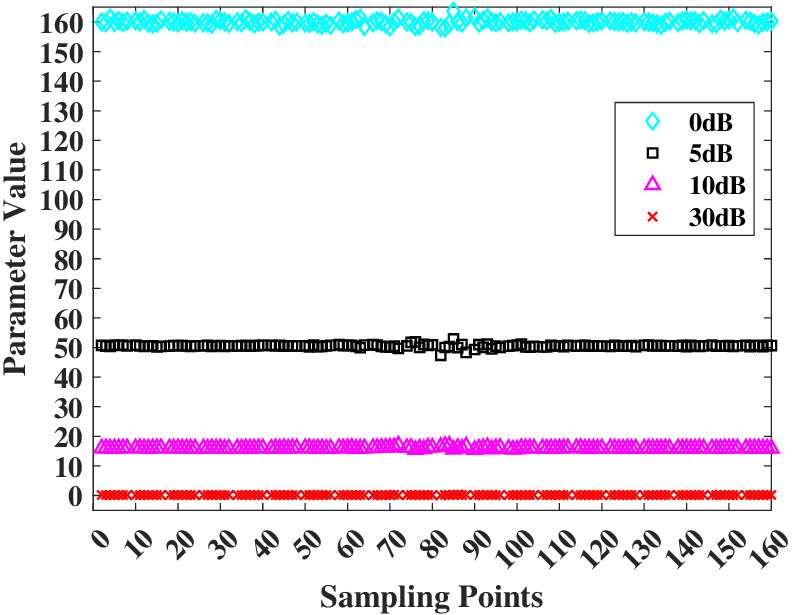}}
        \caption{Estimated parameter values at different SNRs.}
    \label{fig4}
\end{figure}

As can be seen in Fig.~\ref{fig3}, the lower the SNR, the higher the curve of SCPSDs without energy normalization, which is consistent with the analysis about (\ref{eq_12}). In order to verify the reasonableness of (\ref{eq_13}), all the SCPSDs at different SNRs are calculated as mean values separately. Then the SCPSDs at high SNRs, which serve as the ideal RF fingerprint result $\left \| X  \right \| ^{2} \left \| R\!F\!F \right \| ^{2} \left \| X' \right \| ^{2}$, are subtracted. Finally, the value to be observed is obtained by dividing by $\left \| X' \right \| ^{2}$ and comparing whether it is equal to $10^{-0.1 \cdot S\!N\!R}$ or not. Taking device 1 and device 2 as examples, the results of the simulation experiment are shown in Fig.~\ref{fig4}.

The length of a symbol in the simulation experiment is 160, then the corresponding parameter values can be calculated as 160, 50.6, 16 and 0.16 when the SNR is 0~dB, 5~dB, 10~dB and 30~dB respectively. The calculated results match the simulation results in Fig.~\ref{fig4}, proving the validity of (\ref{eq_13}).

\subsection{Anti-noise Algorithm}
From the analysis in the above two subsections, it can be seen that the SCPSD of the received signal at any SNR can be restored to the high SNR case after the normalized energy term processing and RF fingerprint term processing are completed. 

Note that $a=\sum_{i=1}^{L_N}\left \| x_i \right \| ^{2}$ in the normalized energy term in Fig.~\ref{fig2} varies with the device, which is contrary to the classification task in the testing phase. A compromise compensation method is considered, where the average value of the coefficients for all devices is calculated as the coefficient value for the testing phase with the same number of frames per device. Therefore, the equation for normalized energy term compensation for all devices can be expressed as

\begin{equation}
\label{eq_14}
\begin{aligned}
    &\bar{P^2}=\bar{a} \cdot 10^{-\bar{b} \cdot S\!N\!R}+\bar{c}\\
    &\left\{\begin{matrix}
        \bar{a}=\frac{\sum_{i=1}^{N_d}a_i}{N_d}\\
        \bar{b}=\frac{\sum_{i=1}^{N_d}b_i}{N_d}\\
        \bar{c}=\frac{\sum_{i=1}^{N_d}c_i}{N_d}
    \end{matrix}\right.
\end{aligned},
\end{equation}
where $N_d$ denotes the total number of experimental devices.

\section{Experimental Evaluation}

\subsection{Experimental Setup}
\begin{figure}[htbp]
    \centering
    \includegraphics[width=0.4\textwidth]{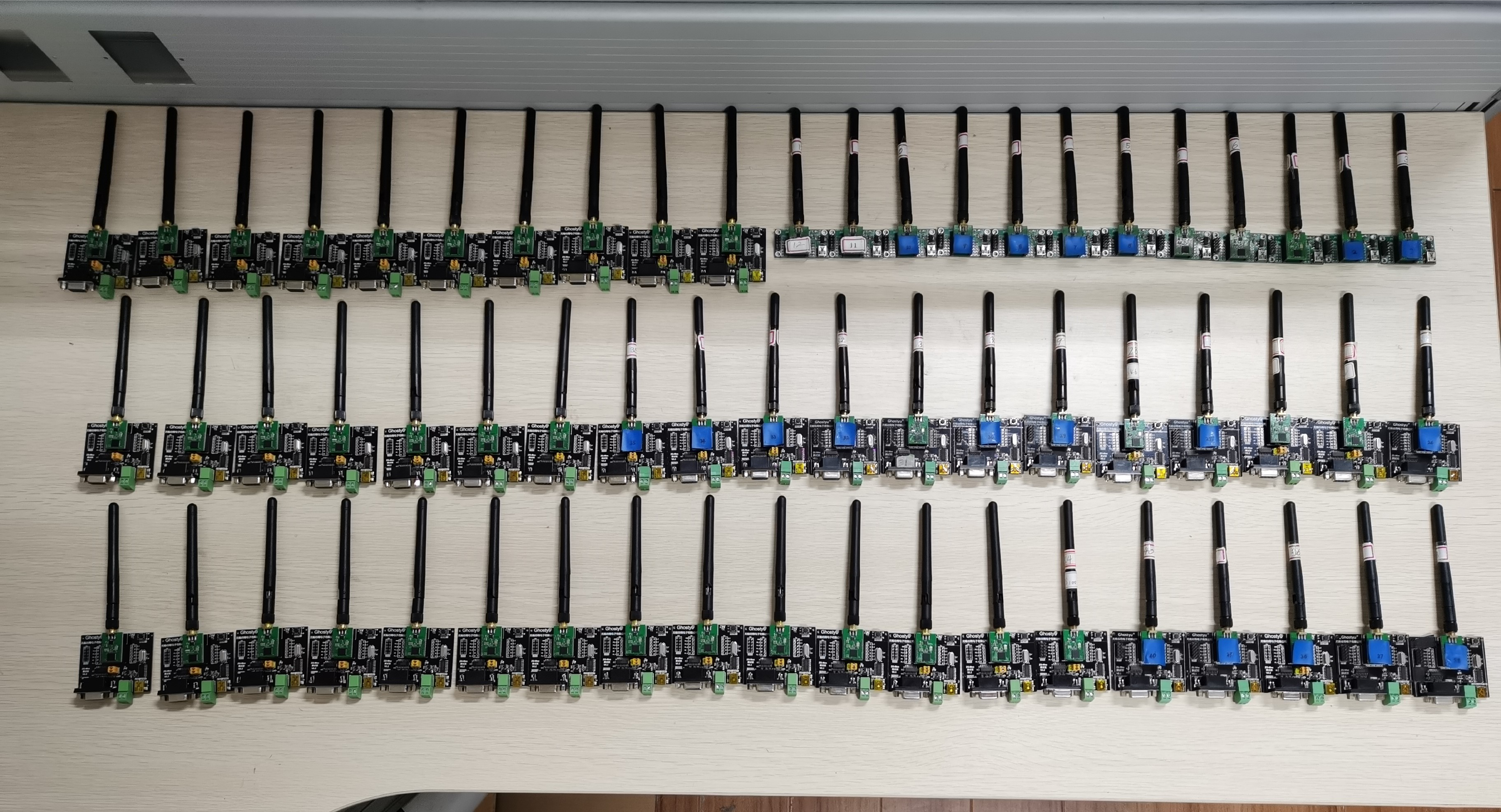}
    \caption{Experimental equipments consisting of 60 off-the-shelf ZigBee devices.}
\label{fig6}
\end{figure}

The experimental devices to be tested are 60 off-the-shelf ZigBee devices based on TI's CC2530 chip, as shown in Fig.~\ref{fig6}. In addition, all devices use Skyworks Solutions corporation's fully integrated RFX2401C front-end module, which provides the RF functionality required by IEEE 802.15.4. The transmitters all work at a center frequency of 2507 MHz with the frame length set to a maximum of 266 and maximum transmit power of 8~dBm. The receiver is a high-performance software defined radio (SDR) platform X310 from Ettus Research Company. It operates at the same center frequency as the transmitters, with a bandwidth of 10 MHz and a sampling rate of 10 MS/s. To achieve a flat fading channel when the signal reaches the receiver, the transmitter is connected to the receiver by an RF coaxial cable \cite{9450821}. A total number of 300 signal frames was sampled for each device, whose corresponding SNRs are all larger than 37~dB. In order to make the training dataset different from the testing dataset, separate AWGN is added to the signal frames in the training phase and in the testing phase.

\subsection{Training Phase}

\subsubsection{Normalized energy item recovery}
Firstly, AWGN is added to these signals to get signal frames at different SNRs. Then the energy of each signal frame is calculated and SNR is estimated to form the coordinate points of $(S\!N\!R,P^2)$. Finally, the coordinate points constituted by all devices are fitted according to (\ref{eq_9}) to obtain the coefficient value of the normalized energy term. The ultimate normalized energy estimation $\bar{P^2}$ is expressed as

\begin{equation}
\label{eq_15}
    \bar{P^2}=0.7937 \cdot 10^{-0.0965 \cdot S\!N\!R}+0.9975.
\end{equation}

\subsubsection{SCPSD extraction}
\begin{figure}[htbp]
    \centering
        \subfloat[Overall view]{
        \label{fig9_1}
        \includegraphics[width=0.23\textwidth]{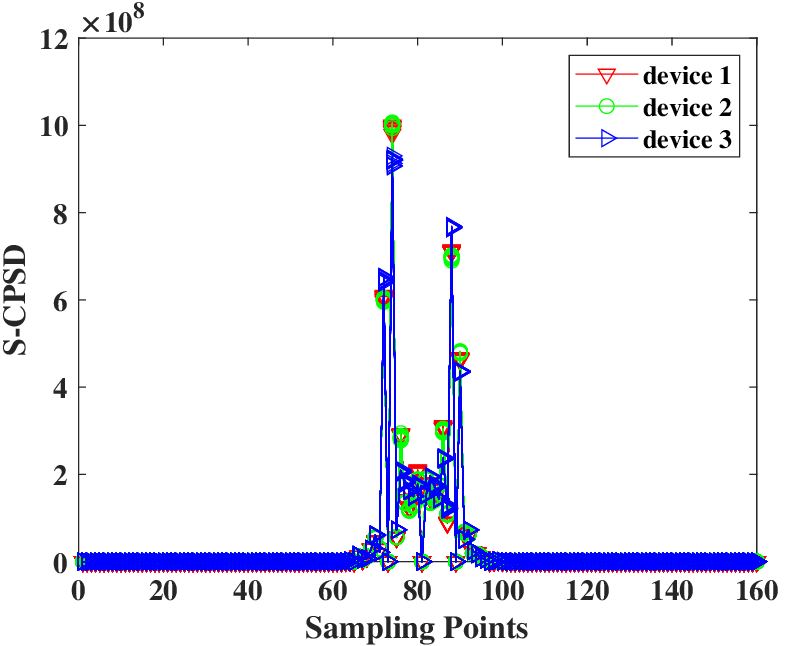}} 
        \subfloat[Local enlargement]{
        \label{fig9_2}
        \includegraphics[width=0.23\textwidth]{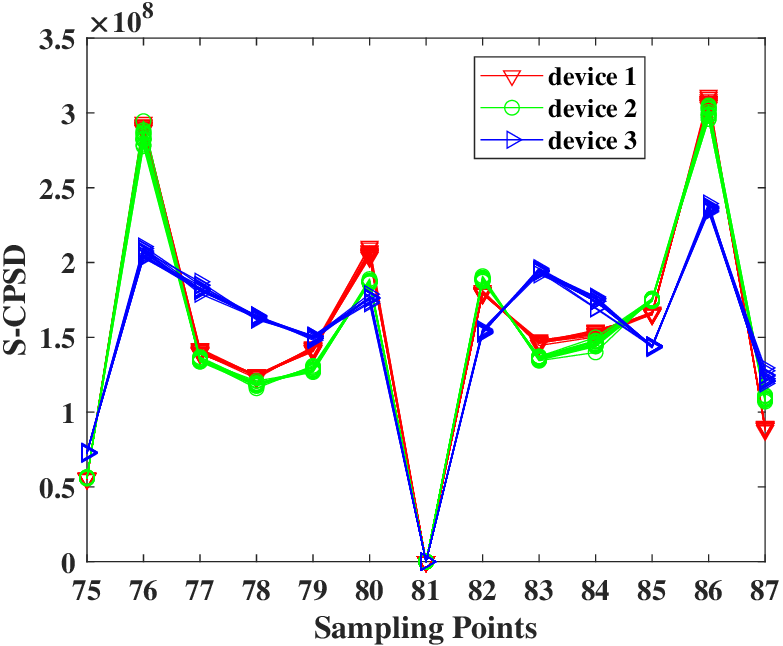}} 
    
        \subfloat[Overall view]{
        \label{fig9_3}
        \includegraphics[width=0.23\textwidth]{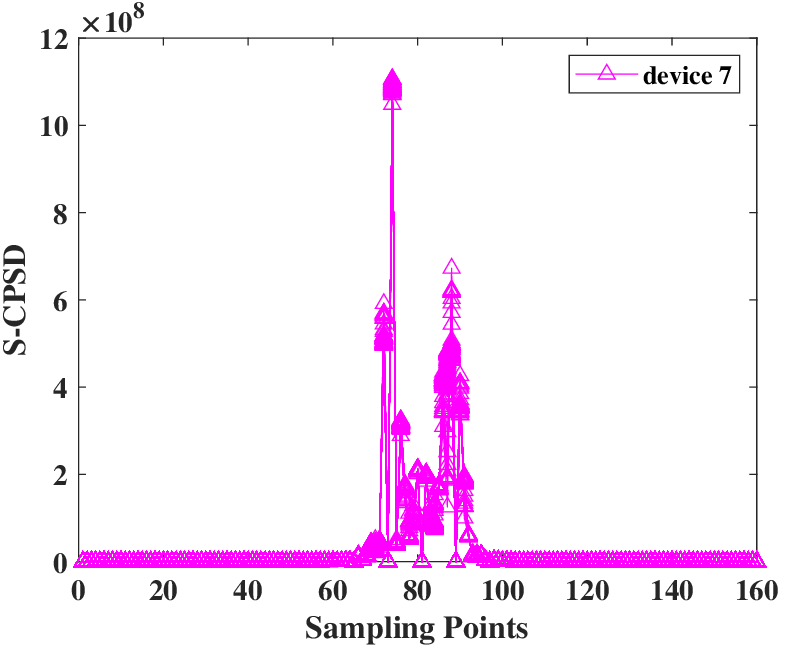}} 
        \subfloat[Local enlargement]{
        \label{fig9_4}
        \includegraphics[width=0.23\textwidth]{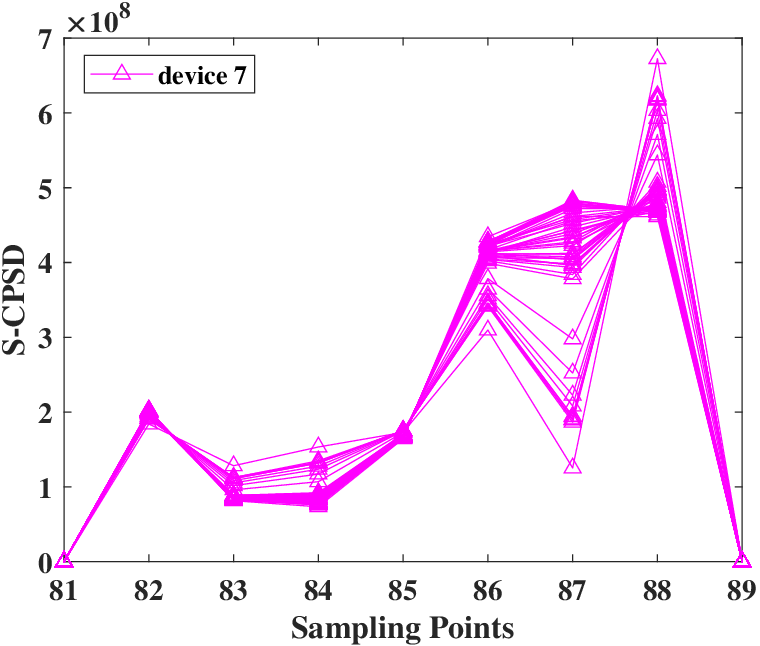}}
    
        \caption{SCPSDs extracted from some devices with the SNR around 40~dB. }
    \label{fig9}
\end{figure}

The SCPSDs of all devices are extracted according to (\ref{eq_5}) and labeled as the training dataset. The SCPSD curves for some of the devices are shown in Fig.~\ref{fig9}. After analyzing the SCPSDs of 60 devices, there were differences in the characteristic curves of various devices. As shown in Fig.~\ref{fig9}\subref{fig9_1} and Fig.~\ref{fig9}\subref{fig9_2}, the respective SCPSD curves for each device are highly aggregated with little fluctuation for device 1, 2 and 3. It can also be seen that the SCPSD curves of device 1 and device 2 are much closer to each other, while device 3 is much more different from them. However, the SCPSD curves of the device 7 in Fig.~\ref{fig9}\subref{fig9_3} and Fig.~\ref{fig9}\subref{fig9_4} fluctuates considerably, reflecting the instability of the output power when the device is operating.

From Fig.~\ref{fig9}, it can be observed that instability in the operation of the actual devices occurs from time to time. As the operating time increases, so does the number of instances in which the device appears to be in an unstable condition. However, this instability is still a manifestation of the device's features. In many studies this is done only by artificially adding RF fingerprints to the ideal signal, which leads to a lack of practicality in the research \cite{10050441}.

\subsubsection{Classifier selection}
\begin{figure}[htbp]
    \centering
        \subfloat[60 devices]{
        \label{fig10_1}
        \includegraphics[width=0.23\textwidth]{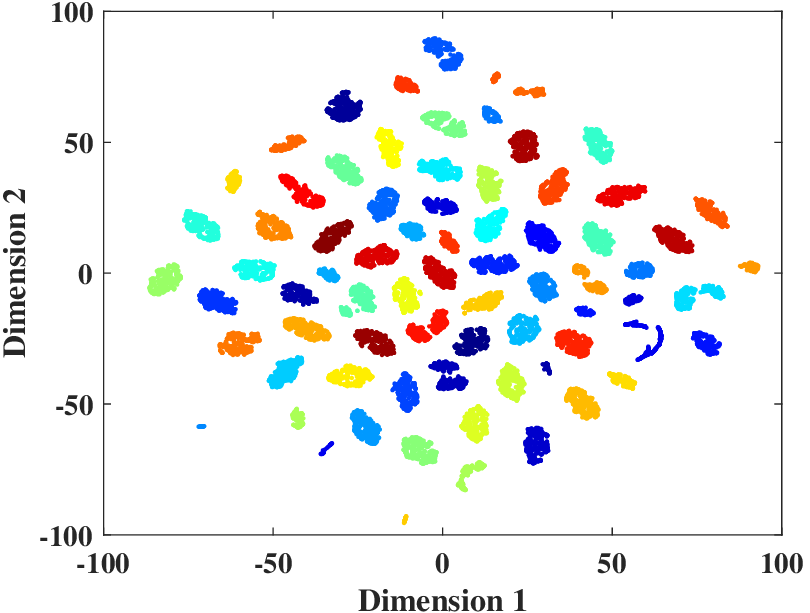}}
        \subfloat[Device 1, 2, 3, 6, 7 and 9]{
        \label{fig10_2}
        \includegraphics[width=0.23\textwidth]{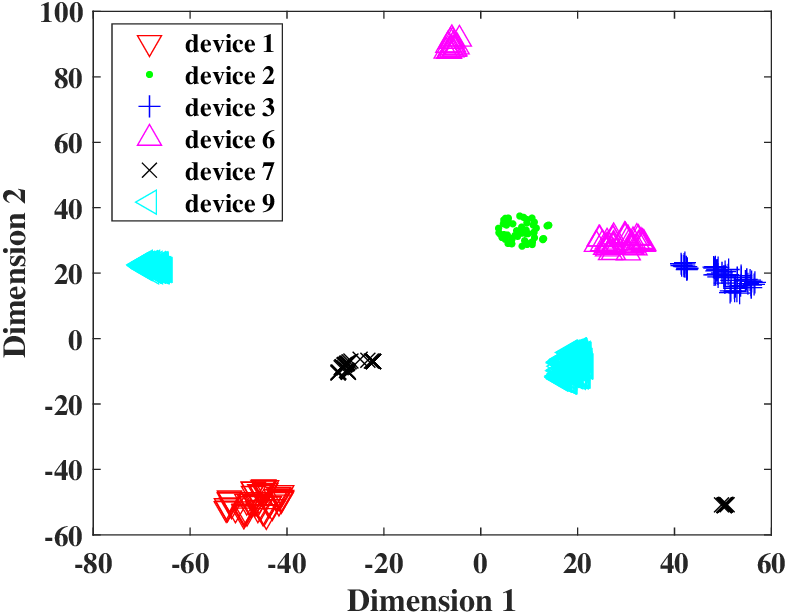}}
        \caption{The visualization results of SCPSDs after dimensionality reduction through t-SNE.}
    \label{fig10}
\end{figure}

To further evaluate the distinguishability of SCPSD on 60 ZigBee devices, t-distributed stochastic neighbor embedding (t-SNE) is utilized to reduce its dimensionality to a two-dimensional plane to visualize the SCPSD features. The result of the 160-dimensional SCPSD after dimensionality reduction is shown in Fig.~\ref{fig10}. In Fig.~\ref{fig10}\subref{fig10_1}, after dimensionality reduction, the SCPSDs of 60 devices show that different classes were isolated from each other, while points of the same class are highly clustered. This shows that the SCPSD features can effectively differentiate the 60 ZigBee devices and achieve high classification accuracy.

Furthermore, it can be seen that the points of device 1, 2 and 3 only cluster in one area in Fig.~\ref{fig10}\subref{fig10_2}, respectively. However, the points of device 7, 8 and 9 are distributed in two areas. This reflects the instability of the device when operating and places demands on the suitability of the classifier. For such a feature distribution, ensemble learning is considered to classify the SCPSDs extracted from 60 ZigBee devices with multi-cluster feature. The random subspace (RS) method of ensemble learning is performed by training any number of weak learners, each of which selects features of any number of dimensions by sampling without replacement. Each weak learner can use different algorithms such as k-nearest neighbors (KNN), discriminant analysis, decision tree and so on. The scores of every weak learner for each class are averaged and the class with the highest average score is categorized. Based on the analysis of the t-SNE visualization results in Fig.~\ref{fig10}, the weak learner uses the KNN algorithm.

To demonstrate the performance of the RS-KNN classifier used in this paper, other machine learning classifiers were used for comparison, including quadratic discriminant analysis (QDA), Support Vector Machine (SVM), KNN, bagged decision tree (BDT), random subspace discriminant analysis (RSDA), multilayer perceptron (MLP). Experiments are conducted using the dataset around 40~dB as the training set and the dataset after adding AWGN as the testing set, whose the results are listed in Table~\ref{tab1}. 

\begin{table}[htbp]
\caption{Accuracy of different classifiers without anti-noise processing.}
\label{tab1}
\centering
\setlength{\tabcolsep}{0.5mm}{
\begin{tabular}{lrrrrrrr}
\hline
\multirow{2}{*}{Dataset}  & \multicolumn{7}{c}{Classifier} \\ \cline{2-8}
  &\quad QDA &\quad SVM &\quad KNN &\quad BDT &\quad RSDA &\quad MLP &\enspace \textbf{RSKNN} \\ \hline
40~dB(train) & 100\% & 100\% & 100\% & 100\% & 100\% & 98.1\% & \textbf{100\%}  \\ 
35~dB(test) & 69.01\% & 13.41\% & 1.67\% & 91.82\% & 99.84\% & 6.12\% & \textbf{100\%}  \\ 
25~dB(test) & 4.63\% & 1.67\% & 1.67\% & 10.46\% & 31.42\% & 3.02\% & \textbf{100\%}  \\ 
20~dB(test) & 2.69\% & 1.67\% & 1.67\% & 9.89\% & 7.19\% & 2.12\% & \textbf{100\%}  \\ 
15~dB(test) & 2.07\% & 1.67\% & 1.67\% & 7.64\% & 2.27\% & 1.81\% & \textbf{91.83\%}  \\ 
10~dB(test) & 1.61\% & 1.67\% & 1.67\% & 3.38\% & 1.67\% & 1.77\% & \textbf{47.06\%}  \\ 
\ \! 5~dB(test) & 1.74\% & 1.67\% & 1.67\% & 3.33\% & 1.67\% & 1.74\% & \textbf{16.06\%}  \\ 
\ \! 0~dB(test) & 1.68\% & 1.67\% & 1.67\% & 3.36\% & 1.67\% & 1.61\% & \textbf{2.88\%}  \\ 
\hline
\end{tabular}}
\end{table}

It can be seen that almost all classifiers can achieve 100\% accuracy on the training set. However, as the SNR decreases, the accuracy of classifiers without ensemble learning decreases much more than those with ensemble learning methods. The results fit well with the previous analysis for the classification task where ensemble learning applies to the RF fingerprints with spatially non-unique distribution. The RSKNN classifier is more accurate than the classifiers being compared at all SNRs, with a significant decrease in accuracy only occurring when the SNR decreases to 10 dB. 

\subsubsection{Testing phase}

To demonstrate the effectiveness of the anti-noise algorithm, the dataset in Table~\ref{tab1} was used for testing. However, since datasets at different SNRs are used in estimating the coefficients of the normalized energy term. Therefore, in order to better test the performance of the algorithm, the AWGN was re-added to the received signals around 40 dB to obtain a new testing dataset. The results are listed in Table~\ref{tab2}. 

\begin{table}[htbp]
\caption{Accuracy of RSKNN classifier with (w/) or without (w/o) anti-noise processing.}
\label{tab2}
\centering
\setlength{\tabcolsep}{1.25mm}{
\begin{tabular}{ccccc}
\hline
\multirow{2}{*}{SNR} & \multicolumn{2}{c}{Old dataset} & \multicolumn{2}{c}{New dataset} \\ \cline{2-5}
        & \  w/o anti-noise & \ w/ anti-noise &\  w/o anti-noise &\ w/ anti-noise \\ \hline
    35dB & 100\% & 100\%  & 100\% & 100\%  \\ 
    25dB & 100\% & 100\%  & 100\% & 100\%  \\ 
    20dB & 100\% & 100\%  & 100\% & 100\%  \\ 
    15dB & 91.83\% & 100\%  & 91.67\% & 100\%  \\ 
    10dB & 47.06\% & 93.83\% & 47.03\% & 93.62\% \\ 
    \ \!   5dB & 16.05\% & 62.52\% & 15.97\% & 62.72\% \\ 
    \ \!   0dB & 2.88\% & 27.60\% & 2.85\% & 27.35\% \\ 
\hline
\end{tabular}}
\end{table}

The training accuracy and testing accuracy can reach 100\% when the SNR is not less than 20 dB. This is reasonable because at high SNR the interference of noise is small and almost negligible. When the SNR is reduced to 15 dB, the accuracy is reduced to about 91\%. However, after anti-noise processing, the accuracy can be increased to 100\% in both old and new dataset. The anti-noise algorithm improves the accuracy by about 46\% for both when the SNR is at 10dB and 5dB, which is a very significant improvement. As the SNR continues to drop to about 0 dB, the boost drops to about 25\%. Overall the anti-noise algorithm is effective, especially for signals with SNR not less than 5 dB.

\section{Conclusion}
In this paper, we first propose the SCPSD feature for RF fingerprint extraction, making the extracted RF fingerprint owns the benefit of data-independent. Subsequently, the specific impact of noise on SCPSD is elaborated through detailed theoretical derivations and simulation experiments. An anti-noise algorithm is proposed by analyzing the various components affected by noise. Addressing the multi-cluster feature problem of RF fingerprints extracted from the real devices, we suggest that utilizing the RSCNN algorithm can achieve better performance compared to other conventional machine learning algorithms. Finally, through practical testing involving 60 off-the-shelf ZigBee devices, the anti-noise algorithm demonstrates the potential for achieving up to approximately 46\% accuracy enhancement under SNR conditions of at least 5 dB.

\bibliography{mybibliography}
\bibliographystyle{IEEEtran}

\end{document}